\RequirePackage{fix-cm}
\documentclass[twocolumn]{svjour3}          
\smartqed  
\usepackage{graphicx}
\usepackage{mathptmx}      
\usepackage{latexsym}
\usepackage[utf8x]{inputenc}
\usepackage{color}
\usepackage{amsmath}
\usepackage{amssymb}
\usepackage{float}
\usepackage{algorithmic}
\usepackage{algorithm}
\usepackage{subfigure}
\usepackage{multirow}
\usepackage{pdfpages}
\usepackage{bm}
\usepackage{bbm}
\usepackage{dsfont}
\usepackage{xspace}
\usepackage{bbm}
\usepackage[authoryear,round]{natbib}
\RequirePackage[colorlinks,citecolor=blue,urlcolor=blue]{hyperref}
\newtheorem{assumption}{Assumption}
\newcommand{\p}{\mathbb{P}}

\newcommand{\E}{\mathbb{E}}

\newcommand{\1}{\mathds{1}}
\newcommand{\iid}{\stackrel{iid}{\sim}}
\newcommand{\NumberOfRuns}{500\xspace}
\newcommand{\NumberOfParticles}{128\xspace}
\newcommand{\TimeHorizon}{1000\xspace}
\newcommand{\stirling}{\genfrac\{\}{0pt}{}}

\begin{document}

\title{Path storage in the particle filter
}


\author{Pierre E. Jacob  \and
Lawrence M. Murray \and
Sylvain Rubenthaler}

\authorrunning{Jacob, Murray \& Rubenthaler} 

\institute{P.E. Jacob \at
    National University of Singapore\\
Department of Statistics \& Applied Probability\\
Block S16, Level 7, 6 Science Drive 2\\
Singapore 117546\\
    \email{pierre.jacob.work@gmail.com}           
    \and
    L.M. Murray \at CSIRO Mathematics, Informatics \& Statistics \\
    Private Bag 5, WA 6913, Australia\\
    \email{lawrence.murray@csiro.au}
    \and
    S. Rubenthaler \at Univ. Nice Sophia Antipolis\\
    Parc Valrose, 06108 Nice cedex 02, France\\
    \email{Sylvain.Rubenthaler@unice.fr}
}

\date{Received: date / Accepted: date}

\maketitle

\begin{abstract}
This article considers the problem of storing the paths generated by a
particle filter and more generally by a sequential Monte Carlo algorithm.
It provides a theoretical result bounding the expected memory
cost by $T + CN\log N$ where $T$ is the time horizon, $N$ is the number of particles and $C$ is a constant, as well as an efficient algorithm to realise this. The theoretical result and
the algorithm are illustrated with numerical experiments.
\end{abstract}

\section{Introduction \label{sec:introduction}}

Consider the problem of filtering in state-space models
\citep{cappe:ryden:2004} defined by 
$X_0 \sim \mu(\cdot)$ and for $t=1, \ldots, T$
\begin{align*}
    &X_t \mid X_{t-1} = x_{t-1}  \sim f(\cdot \mid x_{t-1}),\\
    &Y_t \mid X_{t} = x_{t}  \sim g(\cdot \mid x_{t}).
\end{align*} 
Here $X_{0:T}$ is a hidden Markov chain in some space $\mathcal{X}$ with
initial distribution $\mu$ and transition density $f$. The observations
$Y_{1:T}$ in space $\mathcal{Y}$ are conditionally independent given $x_{1:T}$,
with measurement density $g$.  For any vector $v$, introduce the notation $v_{1:n}
= (v_1, \ldots, v_n)$ and $v^{1:n} = (v^1, \ldots, v^n)$.

We denote by $p_t$ the distribution of the path $X_{0:t}$ given the
observations $y_{1:t}$ available at time $t$, from which the filtering
distribution of $X_t$ given $y_{1:t}$, denoted by $\pi_t$, is a marginal.  The
bootstrap particle filter \citep{gordon:salmon:smith:1993}, described in
Algorithm \ref{algo:bootstrapfilter},  recursively approximates the
distributions $p_{1:T}$, and has borne various other sequential Monte Carlo
methods \citep{doucet:defreitas:gordon:2001,doucet2011tutorial}.
\begin{algorithm}
    \caption{Bootstrap particle filter with $N$ particles}
    \label{algo:bootstrapfilter}
    \begin{tabbing}
        \enspace Draw an initial sample $x_{0}^{1:N} \iid \mu$.\\
        \enspace Set for $k = 1,\ldots, N$, $\bar{x}_{0}^{k} = x_0^k$ and $w_{0}^{k} = 1/N$.\\
        \enspace For $t=1,\ldots,T$\\
        \qquad [resampling] Draw ancestor indices $a_t^{1:N} \sim \mathcal{R}(w_{t-1}^{1:N})$.\\
        \qquad For each $k \in \{1,\ldots,N\}$\\
        \qquad \qquad [transition] Draw a new sample $x_{t}^{k} \sim f(\cdot \mid x_{t-1}^{a_t^k})$.\\
        \qquad \qquad Extend the path $\bar{x}_{0:t}^{k} = (\bar{x}_{0:t-1}^{a_t^k}, x_t^k)$.\\
        \qquad \qquad [weighting] Compute unnormalized weights $\tilde{w}_{t}^{k} = g(y_t \mid x_t^k)$.\\
        \qquad Normalize weights for $k=1,\ldots, N$, $w_{t}^{k} = \tilde{w}_t^k / \sum_{j=1}^N \tilde{w}_t^j$.
    \end{tabbing}
\end{algorithm}
In Algorithm \ref{algo:bootstrapfilter} the resampling step relies on some
distribution $\mathcal{R}$ on $\{1,\ldots, N\}^N$ taking
normalized weights as parameters.

At each time $t$, Algorithm \ref{algo:bootstrapfilter} approximates $p_t$ and $\pi_t$ by
the empirical distributions
\begin{align*}
    p_t^N(dx_{0:t}) &= w_t^1 \delta_{\bar{x}_{0:t}^1}(dx_{0:t}) + \cdots + w_t^N \delta_{\bar{x}_{0:t}^N}(dx_{0:t})\\
    \mbox{and}\quad\pi_t^N(dx_t)   &= w_t^1 \delta_{x_t^1}(dx_t) + \cdots + w_t^N \delta_{x_t^N}(dx_t).
\end{align*}
It has been shown
in \citet{whiteley2011stability,Douc:Moulines:Olsson:2012,vanHandel:2009}, and
in Theorem 7.4.4 in \citet{delMoral:book} that $\pi_t^N$ converges to $\pi_t$
with $N$ under mild conditions on the model laws $(\mu, f, g)$, and that the
Monte Carlo error is constant with respect to $t$.  However it is also
well-known that the path measures $p_{t}^N$, while converging to
$p_t$ with $N$, have a Monte Carlo error typically exploding at least
quadratically with the time $t$ \citep{del2003class,poyiadjis2011particle}.
Indeed the paths quickly coalesce due to the resampling steps, thus providing
a poor approximation of the marginal distributions $p(dx_s | y_{1:t})$ for
large values of $t - s$. In the following we refer to the collection of paths
$\bar{x}_{0:t}^{1:N}$ as the ancestry tree, to each $x^k_s$ (for $k =
1,\ldots,N$ and $s = 0,\ldots,t$) as a node, to each $x^k_t$ more specifically
as a leaf node, and to paths as branches.

\begin{figure*}
 \centering
 \includegraphics[width=0.95\textwidth]{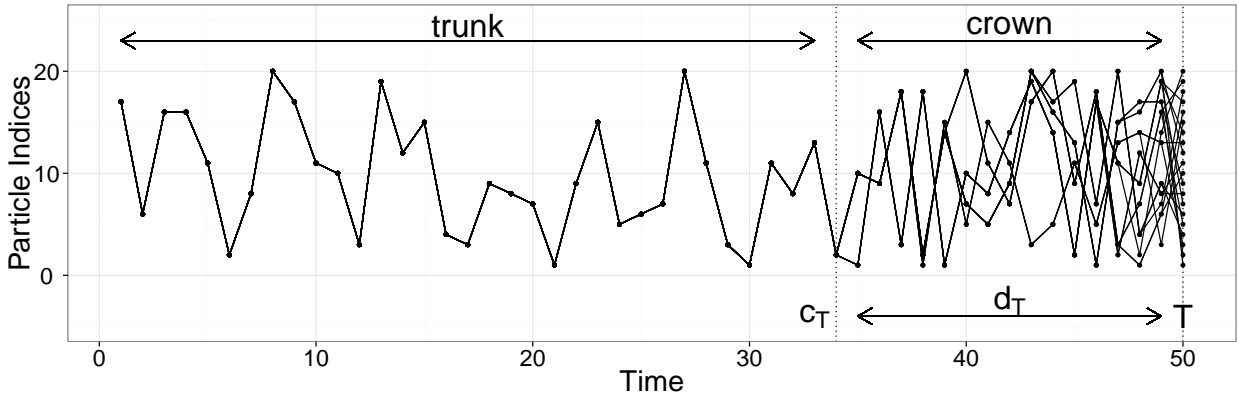}
 \caption{\label{fig:tree} Typical ancestry tree generated by 
 a particle filter using multinomial resampling, with $N = 20$ and $T = 50$.}
\end{figure*}
Figure \ref{fig:tree} might help to visualise the typical shape of the
ancestry tree generated by a particle filter. The time at which all the
branches coalesce, denoted by $c_T$, separates the ``trunk'' made of a unique
branch from $t=0$ to $t=c_T-1$ from the ``crown'' made of all the branches
from $t=c_T$ to $t=T$.  Despite its negative consequence on the estimation of
filtering quantities, the particle degeneracy phenomenon results in crowns of
small sizes, allowing full trees to be stored at low memory cost.  This can be
beneficial whenever full paths of the particle filter are required, such as
for the conditional sequential Monte Carlo and particle Gibbs algorithms first
described in \citet{andrieu:doucet:holenstein:2010}, studied
in \citet{Chopin:Singh:2013}, and used extensively in
\citet{chopin2012smc2} and \citet{lindsten2012ancestor}. Another instance of sequential Monte Carlo method requiring path storage is presented in \citet{wang:jasra:2013}
in the context of computational biology. In the present article algorithms and
results are presented in the filtering terminology, however they immediately extend to any sequential
Monte Carlo method for Feynman-Kac models \citep{delMoral:book}.

In Section \ref{sec:algorithms} we present an efficient algorithm to store ancestry trees
recursively during the run of a particle filter.
In Section \ref{sec:theory} we present new theoretical results bounding the
size of ancestry trees, in order to bound the expected memory requirements of
the storage algorithm.
Finally the theoretical results and the algorithmic performance are tested numerically
in Section \ref{sec:numerics}.

\section{Algorithms\label{sec:algorithms}}

This section introduces a memory-efficient data structure and
associated algorithms for storing only those paths with support at time
$t$. The algorithms are designed for parallel execution, in keeping with the
general parallelisability of other components of sequential Monte Carlo
samplers~\citep{lee2010,murray2013libbi}.

\subsection{Proposed scheme}

Up to time $t$, the particle filter produces particles $x^{1:N}_{1:t}$ and
ancestors $a^{1:N}_{1:t}$. From $a^{1:N}_{1:t}$, offspring counts
$o^{1:N}_{1:t}$ are readily obtained~\citep{murray2013}, where $o^i_t$
represents the number of children at generation $t$ of particle
$x^i_{t-1}$. Let $x^{1:M}_*$ represent $M$ slots in memory for storing
particles. At any time, some of these slots are empty, while others store the
nodes of the tree. Let $a^{1:M}_*$ be an ancestry vector, where $a^i_* = 0$ if
$x^i_*$ is empty or a root node, and otherwise $a^i_* = j$ to indicate that
the particle in $x^j_*$ is the parent of the particle in $x^i_*$. Let
$o^{1:M}_*$ be the offspring vector corresponding to $a^{1:M}_*$, where $o^i_*
= n$ indicates that $x^i_*$ has $n$ children. Finally, let $l^{1:N}_*$ give
the numbers of the $N$ slots in $x^{1:M}_*$ that store the particles of the
youngest generation; these are the leaf nodes of the tree.

Basic operations on the tree are its initialisation, the insertion of a new
generation of particles, and the pruning of older particles to remove those
without a surviving descendent in the youngest generation. These operations
are described in Algorithm \ref{algo:init-insert-prune}. The descriptions
there rely on primitive operations defined in
Algorithm \ref{algo:primitives}. The efficient implementation of such
primitives is well understood in both serial and parallel contexts, so that
they make useful building blocks for the higher-level algorithms.

\begin{algorithm}[tp]

\caption{Parallel algorithms for basic operations on an ancestry tree:
initialising from the first generation of particles, inserting a new
generation of particles, and pruning just before a new generation is
inserted.\label{algo:init-insert-prune}}

    \begin{tabbing}
        \enspace$\textsc{init}(x^{1:N}_0)$\\
        \enspace\qquad For each $i \in \{1,\ldots,M\}$\\
        \enspace\qquad\qquad   $a^i_* \leftarrow 0$\\
        \enspace\qquad\qquad   $o^i_* \leftarrow 0$\\
        \enspace\qquad For each $i \in \{1,\ldots,N\}$\\
        \enspace\qquad\qquad   $x^i_* \leftarrow x^i_0$\\
        \enspace\qquad\qquad   $l^i_* \leftarrow i$
    \end{tabbing}

    \begin{tabbing}
        \enspace $\textsc{insert}(x^{1:N}_t, a^{1:N}_t)$\\
        \enspace\qquad $b_t \leftarrow \textsc{gather}(l_*, a_t)$\\

        \enspace\qquad $z_* \leftarrow \textsc{transform-prefix-sum}(o_*, \mathbf{1}_{\{0\}})$\\
        \enspace\qquad $l_* \leftarrow \textsc{lower-bound}(z_*,(1,\ldots,N))$\\




        \enspace\qquad $a_* \leftarrow \textsc{scatter}(b_t, l_*)$\\
        \enspace\qquad $x_* \leftarrow \textsc{scatter}(x_t, l_*)$
    \end{tabbing}

    \begin{tabbing}
        \enspace $\textsc{prune}(o^{1:N}_t)$\\
        \enspace\qquad $o_* \leftarrow \textsc{scatter}(o_t, l_*)$\\
        \enspace\qquad For each $i \in \{1,\ldots,N\}$\\
        \enspace\qquad\qquad $j \leftarrow l^i_*$\\
        \enspace\qquad\qquad While $j > 0$ and $o^j_* = 0$\\
        \enspace\qquad\qquad\qquad $j \leftarrow a^j_*$\\
        \enspace\qquad\qquad\qquad If $j > 0$\\
        \enspace\qquad\qquad\qquad\qquad $o^j_* \leftarrow o^j_* - 1$
    \end{tabbing}

\end{algorithm}

\begin{algorithm}[tp]
    \caption{Primitives used in pseudocode.\label{algo:primitives}}

    \begin{tabbing}
        \enspace $\textsc{gather}(p^{1:P}, q^{1:Q}) \rightarrow r^{1:Q}$\\
        \enspace\qquad For each $i \in \{1,\ldots,Q\}$\\
        \enspace\qquad\qquad $r^i \leftarrow p^{q^i}$
    \end{tabbing}

    \begin{tabbing}
        \enspace $\textsc{scatter}(p^{1:P}, q^{1:P}) \rightarrow r^{1:Q}$\\
        \enspace\qquad For each $i \in \{1,\ldots,P\}$\\
        \enspace\qquad\qquad $r^{q^i} \leftarrow p^i$
    \end{tabbing}

    \begin{tabbing}
        \enspace $\textsc{transform-prefix-sum}(p^{1:P}, f) \rightarrow r^{1:P}$\\
        \enspace\qquad For each $i \in \{1,\ldots,P\}$\\
        \enspace\qquad\qquad $r^i \leftarrow \sum_{j=1}^i f(p^j)$
    \end{tabbing}

    \begin{tabbing}
        \enspace $\textsc{lower-bound}(p^{1:P}, q^{1:Q}) \rightarrow r^{1:Q}$\\
        \enspace\qquad For each $i \in \{1,\ldots,Q\}$\\
        \enspace\qquad\qquad $r^i \leftarrow \min\{ j: q^i \leq p^j \}$
    \end{tabbing}

\end{algorithm}

To begin, the first of the $M$ empty slots of the tree are initialised with
the first generation of $N$ particles as in the \textsc{init} procedure of
Algorithm \ref{algo:init-insert-prune}. We assume, for now, that $M$ is
sufficiently large to accommodate all subsequent operations on the tree, but
see remarks in Section \ref{sec:algorithms:remarks} below.

Each new generation is inserted as in the \textsc{insert} procedure of
Algorithm \ref{algo:init-insert-prune}.  The procedure searches for nodes with
no offspring in the current generation, and replaces them with the new leaf
nodes.  The vector $z_*$ is introduced, where $z_*^i$ is equal to the number
of nodes between $1$ and $i$ with no offspring. Nodes to replace are then
located by searching for the increments in $z_*$. The new generation is
inserted at these locations.

Finally, the tree is pruned before the insertion of each new generation $t$,
using the \textsc{prune} procedure of Algorithm
\ref{algo:init-insert-prune}. This requires the offspring vector, $o_t$, of
the new generation. The algorithm determines which of the current leaf nodes
have no offspring in the new generation, decrements the offspring counts of
their parent nodes, and proceeds recursively up the tree for cases where the
parent has no remaining offspring either. Each non-leaf node $i$ is considered
pruned if $o^i_* = 0$, and may be overwritten by future calls to
\textsc{insert}.

\subsection{Remarks and improvements\label{sec:algorithms:remarks}}

The \textsc{insert} procedure of Algorithm \ref{algo:init-insert-prune}
assumes that there are at least $N$ free slots in which to place the latest
nodes.  If this is not true, the buffer can be enlarged by allocating a larger
block of memory, copying the contents of the ancestry tree across, and filling
the new regions of the $o_*$ and $a_*$ vectors with zeros. Various heuristics
can be used to set the new size $M$, aiming to reduce fragmentation and the
chance of future increases. Because memory reallocations involve an expensive
copy, it is worth increasing $M$ more than strictly necessary to postpone
additional reallocations. For instance, implementations of the \texttt{C++
Standard Template Library} typically double the storage capacity of a vector
that is extended by just one element, anticipating further extensions. A more
conservative strategy is to start with a value of $M$ equal to a small
multiple of $N$, and enlarge by $N$ slots whenever necessary. Ultimately, we
have not found that the particular enlargement strategy affects execution time
a great deal, particularly since, as in the proceeding theoretical results,
the size of the ancestry tree crown is independent of $T$, so that the need
for reallocations diminishes as $t$ increases.

According to the results of Section \ref{sec:theory}, the expectation of the
size of the tree grows linearly with $T$, but this is only due to the
trunk. The size of the crown is independent of $T$. It may be possible to
improve the algorithms by identifying the nodes along the trunk and storing
them separately, as these nodes will never be overwritten by subsequent
insertions. Under this modified scheme a separate, single growing trunk needs
to be stored but not searched, while the nodes of the crown need to be stored
and searched at every time step. The number of nodes in the crown is of
constant expectation according to Theorem \ref{thm:distance} of
Section \ref{sec:theory}.  Hence this modification induces a scheme of
constant expected computational cost in $T$, which could be relevant in
applications where the time horizon is very long, although there will be
overhead in identifying the trunk. See Fig. \ref{subfig:ComputingTime} in
Section \ref{sec:numerics} for a report on the computational cost of the
proposed method. Memory reallocation is also reduced by storing the trunk
separately.


We establish in Section \ref{sec:theory} that the size of the tree is expected
to be bounded by $T + \Delta_2 N \log N$ for some constant $\Delta_2$. The
size of the data structure, $M$, must be at least as large as this. We assume
that, with a sensible enlargement strategy, it is no more than a constant
factor larger than this, so that its expected memory complexity is
$\mathcal{O}(T + \Delta_2 N \log N)$.

The computational complexity of \textsc{init} is linear in the size of the
data structure, $\mathcal{O}(T + \Delta_2 N \log N)$. A serial implementation
of \textsc{insert} permits a linear prefix sum and search, so
that \textsc{insert} is also $\mathcal{O}(T + \Delta_2 N \log N)$. In parallel, a
linear prefix sum is still achieved~\citep{sengupta2008}, but the search
becomes $N$ binary searches, logarithmic to the size of the data structure;
overall $\mathcal{O}(N \log (T + \Delta_2 N \log N))$.

For \textsc{prune}, consider the best case, where all particles of the
previous generation have an offspring in the new generation. The complexity is
then $\mathcal{O}(N)$: the algorithm operates on each of the $N$ new nodes,
but does not traverse the tree further. Now consider the worst case, where
only one particle of generation $t$ has offspring in the new generation $t +
1$. In this case all but $t$ nodes of the existing tree are pruned, so that
the complexity is $\mathcal{O}(T + \Delta_2 N \log N - t)$ -- linear in the size of the
data structure, and parellelisable.

Finally, the \textsc{transform-prefix-sum} across the full vector $o_*$ in
the \textsc{insert} is redundant. The sum can be truncated once it has reached
$N$, as a sufficient number of free slots have then been found. This is simple
to achieve in the serial case, but it is not obvious how to achieve it in the
parallel case. Heuristic include considering only a subset of $o_*$ at a time
and iterating until a sufficient number of free slots are found, and starting
the cumulative sum after the last slot that was filled in the previous call
to \textsc{insert}. In practice, however, we have observed only negligible
variation in execution times when applying such heuristics, and so have chosen
to present the simplest version here.

\section{Size of the ancestry tree \label{sec:theory}}

\subsection{Results}

From a theoretical point of view, similar random trees have been studied in
population genetics \cite{del2009convergence,mohle2004time} in a setting that
corresponds to a state-space model that assigns equal weights to all paths;
these results do not apply directly here.  In order to bound the expected
number of nodes in an ancestry tree, we first study the distance $d_T = T -
c_T$ between the final time $T$ and the full coalescence time $c_T$ when all
the paths merge. Theorem \ref{thm:distance} proposes a bound on the
expectation of $d_T$, which is independent of $T$ and explicit in $N$.
\begin{assumption}
    \label{as:boundedpotential}
There exists $\epsilon\in[0,1]$ such that for all $y\in\mathcal{Y}$ and for all $x\in\mathcal{X}$
    \[ \sqrt{\epsilon} \leq g(y\mid x) \leq \frac{1}{\sqrt{\epsilon}}.\]
\end{assumption}
\begin{theorem}
   Under Assumption \ref{as:boundedpotential} the distance to the most recent common ancestor $d_T$ satisfies
    \[\E\left[d_T\right] \leq \Delta_1 N \log N\]
    for some $\Delta_1 > 0$, which does not depend $N$ nor $T$.
    \label{thm:distance}
\end{theorem}
  The expected number of nodes in the tree can be bounded explicitly in $N$
and $T$, as in Theorem \ref{thm:numberofnodes}.
\begin{theorem}
   We suppose here that $N\geq 3$. Under Assumption \ref{as:boundedpotential} the number of
    nodes, denoted by $n_{T}$ at time $T$, satisfies
    \[ \E\left[n_{T}\right]\leq T + \Delta_2 N\log N \]
    for some $\Delta_2>0$ that does not depend on $N$ nor $T$.
    \label{thm:numberofnodes}
\end{theorem}
These results quantify the practical difference between storing all the
generated particles (for a deterministic cost of $T\times N$ memory units) and
storing only the surviving particles (for a random cost expected to be bounded
by $T + \Delta_2 N \log N$). 

Assumption \ref{as:boundedpotential} is very strong outside compact spaces, and
for instance does not even cover the linear-Gaussian case, although the
experiments of Section \ref{sec:numerics} indicate that similar results might
hold for non-linear and non-Gaussian cases. The numerical experiments show that
the bound is accurate as a function of $N$, so that even if some inequalities
used in the proofs appear quite crude, the overall result is precise. However
the results do not capture the shape of the tree as a function of $\epsilon$,
which is why we write the constants $\Delta_1$ and $\Delta_2$ without making
their dependency on $\epsilon$ explicit.  Consider for example Theorem
\ref{thm:distance}, where $\Delta_1$ can be defined by $\Delta_1 = 1 +
8/\epsilon$, as will be proven in Section \ref{sec:theory:distance}. If the
bound was sharp as a function of $\epsilon$, it would mean that the time to
full coalescence increases to infinity when $\epsilon$ goes to zero.  However
path degeneracy is expected be more acute for smaller $\epsilon$, since more
variability in the particle weights is then allowed. The dependency on $\epsilon$
in $\Delta_1$ is thus not realistic.
We believe the bounds could in fact be independent of $\epsilon$, by considering $\epsilon = 1$ as the
case corresponding to the largest expectations of $d_T$ and $n_T$; a claim not proven here.

Moreover, the proposed proof relies on the multinomial resampling scheme, while
most practitioners favour more sophisticated schemes
\citep{CarClifFearn,LiuChen,Kitagawa,doucet2011tutorial}.  Figure
\ref{subfig:diffResampling} of Section \ref{sec:numerics} indicates that
similar results hold for these other resampling schemes. There are some obvious
counter-examples, for instance when the measurement density is constant,
leading to equal weights at each step (equivalently $\epsilon = 1$).  Then the
results above hold for multinomial resampling but systematic resampling would
completely obviate the path degeneracy phenomenon. Describing features of ancestry trees
corresponding to general resampling schemes would constitute an interesting avenue of research.

The rest of the section is
devoted to proving Theorem \ref{thm:distance} and Theorem
\ref{thm:numberofnodes}.

\subsection{From non-uniform weights to uniform weights}

We first relate the ancestry process associated with particle filters using
multinomial resampling, with the ancestry process associated with the neutral
case, where all the weights would be equal to $N^{-1}$ at every time step. To
do so we introduce various intermediate processes, starting with the exact multinomial
resampling process denoted by $(A_t)_{t\geq 0}$, then an approximation represented by $(A'_t)_{t\geq 0}$
which provides an almost sure upper bound and eventually a process $(Z_k)_{k\geq 0}$ counting the number of nodes at generation $T-k$
in the neutral case, for a fixed time horizon $T$.

Let us introduce an alternative representation of the multinomial resampling
scheme. For each particle index $j=1,\ldots,N$ at time $t$, draw $V_{t}^{j}$
uniformly in $[0,1]$. If $V_{t}^{j}\leq\epsilon$, draw
$U_{t}^{j}\sim\mathcal{U}([0,1])$ and set $a_{t}^{j}=k$ for $k$ such that
$U_{t}^{j}\in[(k-1)/N,k/N]$. If however $V_{t}^{j}>\epsilon$, draw $a_{t}^{j}$
from $\sum_{1\leq i\leq
N}(w_{t-1}^{i}-\epsilon/N)(1-\epsilon)^{-1}\delta_{i}(\cdot)$.  One can check
that Assumption \ref{as:boundedpotential} ensures that $w_{t-1}^i - \epsilon/N
\geq 0$ for each $1\leq i\leq N$ and that the scheme described above leads to
$\p(a_t^j = k) = w_{t-1}^k$ as in multinomial resampling.  The
alternative representation amounts to a mixture of two steps: one step 
that does not take the weights into account, applied if $V_t^j \leq \epsilon$, and
another step that uses the weights, applied if $V_t^j \geq \epsilon$. This perspective
allows to introduce an approximate resampling scheme represented by the process
$(A'_t)_{t\geq 0}$ described below.

For each time $t$, define $A_{t}:\, j\in\{1,\ldots,N\}\mapsto a_{t}^{j}\in\{1,\ldots,N\}$
and then $A_{t}^{\prime}:\{1,\ldots,N\}\rightarrow\{1,\ldots,N\}$
as follows. For all $j$ in $C_{t}=\{k\in\{1,\ldots,N\}:V_{t}^{k}\leq\epsilon\}$,
set $A_{t}^{\prime}(j)=a_{t}^{j}$. Order the $p$ remaining indices
of the set $\{j\in\{1,\ldots,N\}:V_{j}^{t}>\epsilon\}$ into $\{j_{1}<\dots<j_{p}\}$,
set $A'_{t}(j_{1})=\inf(\{1,\ldots,N\}\backslash A'_{t}(C_{t}))$
and then recursively 
\[
    A'_{t}(j_{k})=\inf(\{1,\ldots,N\}\backslash(A'_{t}(C_{t})\cup\{A'_{t}(j_{1}),\dots,A'_{t}(j_{k-1})\})).
\]
Such a function $A_{t}^{\prime}$ almost surely maps to more unique
values than $A_{t}$ by construction.
It can be seen as a mixture of two steps, as described for $A_t$ above,
but this time neither step relies on the values of the weights.

We write $\lvert u\rvert$ for the cardinal of the image of a function
$u:\{1,\ldots,N\}\rightarrow\{1,\ldots,N\}$. In terms of the functions
$(A_{k})_{k\leq T-1}$, the full coalescence time $c_{T}$ can be
defined as 
\[
c_{T}=\sup\{0\leq k\leq T-1:\mid A_{k}\circ A_{k+1}\circ\dots\circ A_{T-1}\mid=1\},
\]
with the convention $c_T = 0$ in the event $\mid A_{k}\circ A_{k+1}\circ\dots\circ A_{T-1}\mid>1$ for each $0\leq k\leq T-1$, which almost surely satisfies 
$c_{T}\geq c'_{T}$ with 
\[
c'_{T}=\sup\{k\leq T-1:\mid A'_{k}\circ A'_{k+1}\circ\dots\circ A'_{T-1}\mid=1\}.
\]
Indeed since $A'_t$ maps to more unique values than $A_t$ at each time $t$, 
the quantity $\mid A'_k \circ \dots \circ A'_{T-1}\mid$, counting the unique ancestors from 
generation $k$ of the particles at time $T$ when using the resampling scheme $A'$, is almost surely
larger than $\mid A_k \circ \dots \circ A_{T-1}\mid$ for any $k$, and hence
it takes longer to reach the full coalescence time when using $A'$ compared to $A$.

Following \cite{del2009convergence}, Section 4 and \cite{mohle2004time},
the sequence $(K_{k})_{k\geq0}=(\mid A'_{T-k}\circ\dots\circ A'_{T-1}\mid)_{k\geq0}$
is a Markov chain in the filtration $(\mathcal{F}_{k})_{k\geq1}$
with 
\[
\mathcal{F}_{k}=\sigma(V_{r}^{1:N},U_{r}^{1:N})_{T-k\leq r\leq T-1},
\]
with the convention $K_{0}=N$. For all $k\geq0$, $q\in\{1,\ldots,N\}$
and $p<q$ its transition law verifies 
\begin{align}
    & \p(K_{k+1}=p\mid K_{k}=q)\label{eq:transitionofK} \\
 &=\sum_{q'=q-p+1}^{q}\binom{q}{q'}\epsilon^{q'}(1-\epsilon)^{q-q'}\stirling{q'}{q'-q+p}\frac{(N)_{q'-q+p}}{N^{q'}}\nonumber \\
 & \text{and}\quad p_{N,q}=\p(K_{k+1}=q\mid K_{k}=q) \nonumber\\
 &=\sum_{q'=0}^{q}\binom{q}{q'}\epsilon^{q'}(1-\epsilon)^{q-q'}\frac{(N)_{q'}}{N^{q'}}\label{eq:definition:pNq}
\end{align}
where $\stirling{q}{p}$ is the Stirling number of the second kind
giving the number of ways of partitioning the set $\{1,\ldots,q\}$
into $p$ non empty blocks and where $(N)_{p}=N!/(N-p)!$. Note that
Eq. \eqref{eq:definition:pNq} is a special case of Eq. \eqref{eq:transitionofK}.

Let us give more details on Eq. \eqref{eq:transitionofK} and \eqref{eq:definition:pNq}. First consider
the expression of $p_{N,q}$. The index $q'$ represents the number
of particles associated with realisations of $V_{T-k-1}$ being less
than $\epsilon$. Hence it is the number of particles of step $T-k-1$
for which the ancestor $A'_{T-k-1}$ was chosen according to the uniform
distribution on $\{1,\ldots,N\}$; the remaining $q-q'$ ancestors
are chosen deterministically; see the definition of $(A'_{t})$. The
term $\tbinom{q}{q'}\epsilon^{q'}(1-\epsilon)^{q-q'}$ corresponds
to the probability of obtaining $q'$ uniform draws of $V_{T-k-1}$
with values less than $\epsilon$ among $q$ particles at time $T-k$.
The term $(N)_{q'}/N^{q'}$ corresponds to the probability of these
$q'$ ancestors, drawn uniformly on $\{1,\ldots,N\}$, landing on
$q'$ unique values. Now consider the probability $\p(K_{k+1}=p\mid K_{k}=q)$
for some $p<q$. 
For $K_{k}$ to fall from $q$ to $p$ at the next
step, $q-p$ unique particles must disappear; since particles
corresponding to $V_{T-k-1} > \epsilon$ do not disappear, there must be
at least $q-p+1$ particles corresponding to $V_{T-k-1} \leq \epsilon$.
Hence the index $q'$, still representing the number of particles with
realisations of $V_{T-k-1}$ less than $\epsilon$, now starts at $q-p+1$. The
binomial term is similar to the case where $p=q$. Among the $q'$ particles with
realisations of $V_{T-k-1}$ less than $\epsilon$, $p'=p-(q-q')$ of them must
choose unique ancestors and the other $q-p$ must coalesce. The Stirling number
$\stirling{q'}{p'}$ indeed counts the number of partitions (groups of particles
that will coalesce) of $\{1,\ldots,q'\}$ in $p'$ non-empty blocks (each
corresponding to a unique ancestor).

Note that conditional upon $K_{k}=q$ there can be any number $I\in\{1,\ldots,q\}$
of variables $V^{1:q}$ falling under $\epsilon$. We can write 
$\E[K_{k+1}\mid K_{k}=q]$ as 
    \[ \sum_{i=0}^{q}\binom{q}{i}\epsilon^{i}(1-\epsilon)^{q-i}\E\left[K_{k+1}\mid K_{k}=q,I=i\right].  \]
We now focus on $\E\left[K_{k+1}\mid K_{k}=q,I=i\right]$, the expected
number of ancestors of $q$ different particles, given that $i$ of
them choose their ancestors uniformly in $\left\{ 1,\ldots,N\right\} $
and that $q-i$ have a unique ancestor. Of course the difficulty comes
from the random component, \emph{id est} the $i$ particles that choose their ancestors uniformly. Introduce
the process $\left(Z_{k}\right)_{k\geq0}$ on $\mathbb{N}$ corresponding
to the number of ancestors in a scheme using only those uniform selections,
which is equivalent to a multinomial resampling scheme with uniform weights.
More formally the transition of $(Z_{k})_{k\geq0}$ satisfies 
\begin{equation}
    \p\left(Z_{k+1}=p\mid Z_{k}=q\right)=\stirling{q}{p}\frac{(N)_{p}}{N^{q}}, \label{eq:transitionZ}
\end{equation}
following the same reasoning as for the transition probabilities of $(K_k)_{k\geq 0}$.
The initial distribution of $Z_0$ is not used in the following hence we do not need to specify it.
The link between $\left(Z_{k}\right)_{k\geq0}$ and $\left(K_{k}\right)_{k\geq0}$
is explicitly given by 
\[
\E\left[K_{k+1}\mid K_{k}=q,I=i\right]=(q-i)+\E\left[Z_{k+1}\mid Z_{k}=i\right]
\]
so that we have 
\begin{align}
&\E\left[K_{k+1}\mid K_{k}=q\right] \nonumber \\
& =q\left(1-\epsilon\right)+\sum_{i=0}^{q}\binom{q}{i}\epsilon^{i}(1-\epsilon)^{q-i}\E[Z_{k+1}\mid Z_{k}=i].\label{eq:linkKandZ}
\end{align}

Note that the process $(Z_{k})_{k\geq0}$ is not used in the proof
of Theorem \ref{thm:distance}, where we start from $(K_{k})_{k\geq0}$
again, but is pivotal for the proof of Theorem \ref{thm:numberofnodes}.

\subsection{Distance to the most recent common ancestor\label{sec:theory:distance}}

We start with the proof of Theorem \ref{thm:distance}. We define
a Markov chain $(L_{k})_{k\geq0}$ on $\mathbb{N}$ such that $L_{0}=N$
and its transition satisfies
\[
\p(L_{k+1}=q-1\mid L_{k}=q)=\sum_{p<q}\p(K_{k+1}=p\mid K_{k}=q)
\]
and thus for all $k\geq0$ and $p\leq q$ 
\[
\p(L_{k+1}=p\mid L_{k}=q)=\begin{cases}
p_{N,q} & \,\mbox{if }p=q\,,\\
1-p_{N,q} & \mbox{ if }p=q-1\, ,
\end{cases}
\]
where $p_{N,q}$ is defined in Eq. \eqref{eq:definition:pNq}.
In addition we couple $(L_{k})_{k\geq0}$ and $(K_{k})_{k\geq0}$ by assuming
\begin{itemize}
\item $\left[L_{k}=K_{k}\right]$$\Rightarrow$$\left[L_{k+1}<L_{k}\Leftrightarrow K_{k+1}<K_{k}\right]$
(if the two chains are at the same point, then if one of them decreases,
the other one decreases too)
\item $[L_{k}\neq K_{k}]\Rightarrow$$K_{k+1}$and $L_{k+1}$ are independent,
conditionally upon $L_{k}$, $K_{k}$.
\end{itemize}
By construction $L_{k}\geq K_{k}$ for all $k\geq0$ almost surely.
Hence $c'_{T}\geq T-\mathcal{D}_{T}$ with $\mathcal{D}_{T}=\inf\{k\geq1:L_{k}=1\}$
and thus $d_{T}=T-c_{T}\leq T-c'_{T}\leq\mathcal{D}_{T}$ almost surely.

For $q=2,\ldots,N$ denote by $J_{q}^{(N)}$ the time required for
$(L_{k})_{k\geq0}$ to jump from $q$ to $q-1$. Each $J_{q}^{(N)}$
follows a geometric law with parameter $(1-p_{N,q})$ and $\mathcal{D}_{T}=\sum_{q=2}^{N}J_{q}^{(N)}$,
so that \ensuremath{\E[\mathcal{D}_{T}]=\sum_{q=2}^{N}(1-p_{N,q})^{-1}}.
To conclude, we manipulate this sum as follows. 
For any $k=1,\ldots,N$ a crude bound on $(N)_k/N^k$ is given by $\exp\{-k/2N\}$, 
from which we obtain 
\[p_{N,q} \leq \left(1-\epsilon(1-e^{-1/2N})\right)^{q}.\]
We have, for all $N$, $(8N)^{-1}\leq1-\exp\{-1/2N\}$ and
for all $x\geq1$ and $\varepsilon\in(0,1)$,  $(1-\varepsilon/x)^{x}\leq\exp(-\varepsilon)$;
combining these inequalities we obtain
\[\E[\mathcal{D}_{T}] \leq \sum_{q=2}^{N}(1-\alpha^{q/N})^{-1}\]
where $\alpha=\exp(-\epsilon/8)$.
We can now bound this series by expanding $\alpha^{q/N} = \exp\{(q/N) \log \alpha\}$ into an alternating series
and by bounding the alternating series always by one of its partial sums:
\begin{eqnarray*}
& & \sum_{q=2}^{N}(1-\alpha^{q/N})^{-1} \\
 & \leq & \sum_{q=2}^{N}\left(\frac{q}{N}(-\log\alpha)-\frac{1}{2!}\left(\frac{q}{N}\right)^{2}(\log\alpha)^{2}\right)^{-1}\\
 & \leq & -\frac{N}{\log\alpha}\log N+(N-1) \leq  \left(1+\frac{8}{\epsilon}\right)N\log N,
\end{eqnarray*}
which concludes the proof of Theorem \ref{thm:distance}.

Note that bounding $(K_k)_{k\geq 0}$ by $(L_k)_{k\geq 0}$ almost surely seems very crude, since
$K_k$ can possibly jump from $q$ to $p << q$ in one step whereas $L_k$ can only jump
from $q$ to $q-1$. However the time to coalescence is mostly dominated by the final jumps, 
because the probabilities $\p(K_{k+1} = p\mid K_k = q)$ are close to $0$ when $N$ is large compared to $q$ and $p<q$.
In other words after a few time steps, $q$ is small compared to $N$ and then $(K_k)_{k\geq 0}$ 
mostly jumps from $q$ to $q-1$ if it jumps at all, so that $(L_k)$ provides an accurate bounding process.
The additional approximations used to bound $p_{N,q}$ and thus $\E[\mathcal{D}_T]$ are also to be considered
in the regime of small $q$ compared to $N$, where they prove accurate enough to obtain the desired result in $N\log N$.

\subsection{Number of nodes in the ancestry tree\label{sec:theory:numberofnodes}}

We now proceed to the proof of Theorem \ref{thm:numberofnodes}. Denote
by $m_{T}$ the number of nodes in the crown. The bound on $d_{T}$
from Theorem \ref{thm:distance} gives a first crude bound 
\[
\E[m_{T}]\leq \Delta_1 N^{2}\log N
\]
which is obtained by bounding the size of every generation in the
crown by $N$. However we can obtain a better bound, in $N\log N$,
by the following arguments.

The process $\left(K_{k}\right)_{k\geq0}$ was already introduced
to bound $\E[d_{T}]$ but we can naturally use it to bound $\E[m_{T}]$
since 
\[
m_{T}\leq\sum_{k=0}^{\tau_{T}}K_{k}
\]
almost surely, where $\tau_{T}=\inf\left\{ k\leq T:K_{k}=1\right\} $;
note that $\tau_{T}=T-c'_{T}$.
To bound $\E[K_k]$ we use the chain $(Z_k)_{k\geq 0}$ defined by Eq. \eqref{eq:transitionZ}.
By definition of $(Z_k)_{k\geq 0}$ and denoting by $(C_j)_{j=1}^N$ independent uniform variables in $\{1, \ldots, N\}$, we have
\begin{align}
    \E[Z_{k+1}\mid Z_{k}=q] &= \E[\sum_{j=1}^N \1_{\{\exists k\in\{1,\ldots,q\}: \, C_k = j\}} ]\nonumber\\
    &= N - \sum_{j=1}^N \E[\1_{\{\forall k\in\{1,\ldots,q\}: \, C_k \neq j\}} ]\nonumber\\
    &= N - N \left(1 - \frac{1}{N}\right)^q  \label{eq:CondExpectOfZ}
\end{align}
which, using Eq \eqref{eq:linkKandZ}, implies 
\[
\E\left[K_{k+1}\mid K_{k}=q\right]=q(1-\epsilon)+N\left(1-\left(1-\frac{\epsilon}{N}\right)^{q}\right).
\]
By expanding $(1 - (1 - \epsilon/N)^q)$ into its alternating series and bounding the series by its third partial sum,
we obtain 
\begin{align*}
\E\left[K_{k+1}\mid K_{k}=q\right]&\leq q-\frac{\epsilon^{2}}{2N}q(q-1)+\frac{\epsilon^{3}}{6N^{2}}q(q-1)(q-2).
\end{align*}
Now for $x\in[1,N]$ define the function $g_{N,\epsilon}$ by: 
\begin{equation}
g_{N,\epsilon}(x)=x-\frac{\epsilon^{2}}{2N}x(x-1)+\frac{\epsilon^{3}}{6N^{2}}x(x-1)(x-2).
    \label{eq:definition:gNepsilon}
\end{equation}
Noting that $g_{N,\epsilon}$ is concave and using Jensen's inequality,
we obtain
\begin{equation}
\E[K_{k+1}]=\E[\E[K_{k+1}\mid K_{k}]]\leq g_{N,\epsilon}(\E[K_{k}]).\label{eq:ExpectOfK}
\end{equation}
Introduce the sequence $u_{0}=N$, and $u_{n+1}=g_{N,\epsilon}(u_{n})$
for $n>0$. By the above inequality and because $g_{N,\epsilon}$ is nondecreasing, we have $\E(K_k)\leq u_k$ for all $k$. We can finally bound the expected number of nodes in the
crown as follows 
\begin{align}
\E\left[\sum_{k=0}^{\tau_{T}}K_{k}\right] 
 & =\E\left[\E\left[\sum_{k=0}^{\tau_{T}}(K_{k}-1)\right]+\tau_{T}\right]\nonumber \\
 & \leq\sum_{k=0}^{\infty}(u_{k}-1)+ \Delta_1 N\log N \label{eq:boundonsum}
\end{align}
using Theorem \ref{thm:distance} to bound $\E[\tau_T]$.
We use the following technical lemma to bound $\sum_{k=0}^{\infty}(u_{k}-1)$.

\begin{lemma} Let $N\in\mathbb{N}$, $N\geq 6$ and $\epsilon\in(0,1)$. Consider
    the sequence $(u_{k})_{k\geq0}$ such that $u_{0}=N$ and for $k\geq1$
    \[ u_{k}=u_{k-1}-\frac{\epsilon^{2}}{2N}u_{k-1}(u_{k-1}-1)+\frac{\epsilon^{3}}{6N^{2}}u_{k-1}(u_{k-1}-1)(u_{k-1}-2).  \]
    Then there exists $C>0$ independent of $N$ such that 
    \[ \sum_{k=0}^{\infty}(u_{k}-1)\leq CN\log N. \]
    \label{lemma:nastyseries} 
\end{lemma} 
The proof of Lemma \ref{lemma:nastyseries}, based on elementary real analysis, is given in Appendix \ref{proof:nastyseries}. Using Lemma \ref{lemma:nastyseries}
and Eq. \eqref{eq:boundonsum} we obtain Theorem \ref{thm:numberofnodes}
with $\Delta_{2}=C+\Delta_1$.

\section{Numerical experiments\label{sec:numerics}}

This section provides numerical experiments to illustrate the results of
Section \ref{sec:theory} and the efficiency of the algorithms presented in
Section \ref{sec:algorithms}. The results summarise $K = \NumberOfRuns$
independent runs, using $N = \NumberOfParticles$ particles and
$T \leq \TimeHorizon$ time steps. For each run, a new synthetic dataset is
generated and a different random seed is used.  The default resampling scheme
is the multinomial scheme, applied at every time step. The algorithms of
Section \ref{sec:algorithms} have been implemented in LibBi~\cite[][\url{www.libbi.org}]{murray2013libbi}, which is used for the numerical
results here.

We use the Phytoplankton-Zooplankton (PZ) model described
in \cite{jones2010bayesian} and \cite{murray2012collapsed}.  Concentrations
of phytoplankton $(P_t)$ and zooplankton $(Z_t)$, along with the stochastic
growth rate of phytoplankton $(\alpha_t)$, constitute the hidden state. The
state follows the continuous-time dynamics $dP/dt = \alpha_t P - c PZ$ and
$dZ/dt = ec PZ - m_l Z - m_q Z^2$, with
$\alpha_t \sim \mathcal{N}(\mu, \sigma^2)$ drawn at every integer time
$t$. The initial conditions are $\log P_0 \sim \mathcal{N}(\log(2), 0.2)$,
$\log Z_0 \sim \mathcal{N}(\log(2), 0.1)$. The observations $(Y_t)$
measure $(P_t)$ with additive log-normal noise, that is $\log Y_t \sim \mathcal{N}(\log P_t, \sigma_y)$. The parameters are set to $\mu =
0.4$, $\sigma = 0.2$, $c = 0.25$, $e = 0.3$, $m_l = m_q = 0.1$ and $\sigma_y = 0.2$.

\begin{figure}
 \centering
\subfigure[Adjusted number of nodes $\tilde{n}_T = (n_T - T) / N$ against $T$ for various $N$]{\label{subfig:diffN}
\includegraphics[width=0.45\textwidth]{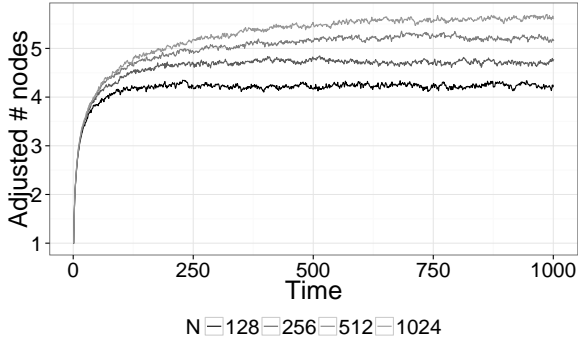}}
\subfigure[Adjusted number of nodes $\tilde{n}_T = (n_T - T) / N$ against $N$ (log-scale) for various $T$]{\label{subfig:diffN2}
\includegraphics[width=0.45\textwidth]{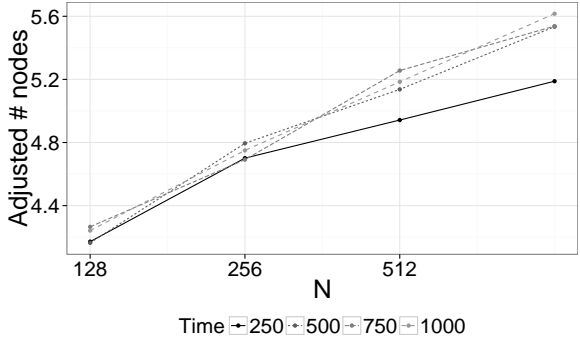}}
 \caption{\label{fig:diffN} Adjusted number 
 of nodes $\tilde{n}_T = (n_T - T) / N$ versus $T$ for various $N$ (top), and versus $N$ for various times $T$ (bottom), for the PZ model.}
\end{figure}

\begin{figure}
 \centering
\subfigure[Adjusted number of nodes $\tilde{n}_T = (n_T - T) / N$ against $T$ for various resampling schemes]{\label{subfig:diffResampling}
\includegraphics[width=0.45\textwidth]{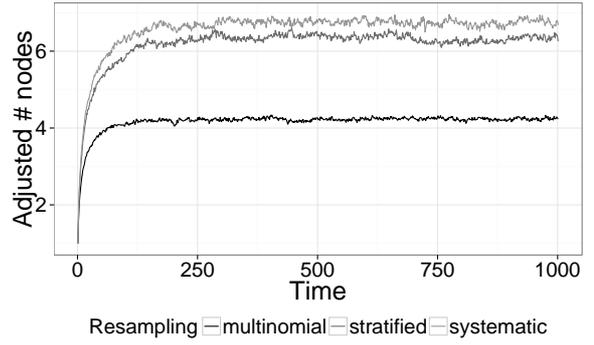}}
\subfigure[Computing time (in microseconds) of the path keeping algorithm against $N$ for various $T$]{\label{subfig:ComputingTime}
\includegraphics[width=0.45\textwidth]{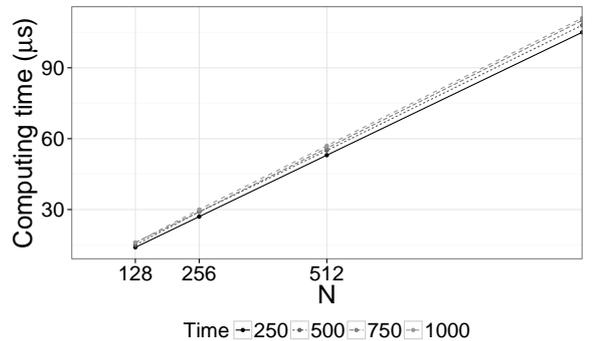}}
 \caption{\label{fig:CTandRes}Impact of the resampling scheme on the number of nodes (top) and computing time of the path keeping algorithm
 for various $N$ and $T$ (bottom), for the PZ model.}
\end{figure}

Lemma \ref{thm:numberofnodes} is illustrated by plots of the adjusted number
of nodes defined by $\tilde{n}_T = (n_T - T) / N$ for various $N$ against $T$
on Fig. \ref{subfig:diffN} and for various $T$ against $N$ on
Fig. \ref{subfig:diffN2}. The quantity is averaged over $K$ independent runs. According to the lemma $\tilde{n}_T$ should be
uniformly bounded as a function of $T$ and should grow logarithmically as a function of $N$;
this is confirmed by the graphs.  Figure \ref{subfig:diffResampling} shows that a similar
behaviour is expected for other resampling schemes such as stratified and
systematic, only with a different value for $\Delta_2$.

To illustrate the efficiency of the procedures presented in
Section \ref{sec:algorithms}, Fig. \ref{subfig:ComputingTime} shows the
combined time taken to execute the pruning and insertion algorithms at each
time step, for various $T$ and $N$.  The results suggest that the
computational cost is not greatly influenced by $T$, and close to linear with
respect to $N$: evidence of a practical implementation with comparable
complexity to the particle filter itself.

\section{Conclusion}

We have presented a bound on the expected number of nodes in
the ancestry tree produced by particle filters.  
The numerical experiments of Section \ref{sec:numerics} indicate that the
result is accurate, even outside the scope of the assumptions made in the theoretical study,
and that the proposed algorithm to store the tree is computationally efficient.

\appendix

\section{Proof of Lemma \ref{lemma:nastyseries}\label{proof:nastyseries}}

Let $N\in\mathbb{N}$ and $\epsilon\in(0,1)$, define $(u_{k})_{k\geq0}$
as in the statement of the lemma and define $g_{N,\epsilon}$ as in Eq.  \eqref{eq:definition:gNepsilon}.
We are interested in $\sum_{k\geq0}(u_{k}-1)$. Note first that $g_{N,\epsilon}$
is contracting and is such that $g_{N,\epsilon}(1)=1$, so that $u_{k}$
goes to $1$ using Banach fixed-point theorem. The contraction coefficient
of $g_{N,\epsilon}$ can be bounded by 
\begin{align*}
\sup_{x}\lvert g_{N,\epsilon}'(x)\rvert & \leq g_{N,\epsilon}'(1)=1-\frac{\epsilon^{2}}{2N}<1,
\end{align*}
however this contraction coefficient depends on $N$ and a direct
use of it yields a bound on $\sum_{k\geq0}(u_{k}-1)$ that is not
in $N\log N$.

Note also that even though $u_{k}$ goes to $1$, we can focus on
the partial sum $\sum_{k=0}^{\sigma_{2}}(u_{k}-1)$ where $\sigma_{2}=\inf\{k:u_{k}\leq2\}$,
because $\sum_{k=\sigma_{2}}^{\infty}(u_{k}-1)$ is essentially bounded
by $N$. Indeed note that for $1\leq u\leq2$ we have 
$(\epsilon^{3}/ 6N^{2})u(u-1)(u-2)\leq0$
so that 
\begin{align*}
u_{k}-1 & \leq u_{k-1}-1-\frac{\epsilon^{2}}{2N}u_{k-1}(u_{k-1}-1) \leq(u_{k-1}-1)(1-\frac{\epsilon^{2}}{2N}), 
\end{align*}
hence $\sum_{k=\sigma_{2}}^{\infty}(u_{k}-1)\leq (2N/ \epsilon^2)$.
Therefore we can focus on bounding $\sum_{k=0}^{\sigma_{2}}(u_{k}-1)$
by $N\log N$. Let us split this sum into partial sums, where the
first partial sum is over indices $k$ such that $N/2\leq u_{k}\leq N$,
the second is over indices $k$ such that $N/4\leq u_{k}\leq N/2$,
etc. More formally, we introduce $(k_{j})_{j=0}^{J}$ such that $k_{0}=0$,
$k_{1}=\inf\{k:u_{k}\leq N/2\}$, \ldots{}, $k_{j}=\inf\{k:u_{k}\leq N/2^{j}\}$,
up to $k_{J}=\inf\{k:u_{k}\leq N/2^{J}\}$ where $J$ is such that
$N/2^{J}\leq2$, or equivalently $\log N/\log2-1\leq J$.
For instance we take $J=\lfloor\log N/\log2\rfloor$. Thus we have split
$\sum_{k=0}^{\sigma_{2}}(u_{k}-1)$ into $J$ partial sums of the
form $\sum_{k=k_{j}}^{k_{j+1}-1}(u_{k}-1)$
and we are now going to bound each of these partial sum by the same
quantity $C(\epsilon)N$ for some $C(\epsilon)$ that depends only
on $\epsilon$.

To do so, we consider the time needed by $(u_{k})_{k\geq0}$ to decrease from
a value $N/m_{j}$ to a value $N/m_{j+1}$, with $m_{j+1}>m_{j}$;
we will later take $m_{j}=2^{j}$ and $m_{j+1}=2^{j+1}$. Note that
for any $m$ we have 
\begin{align*}
g_{N,\epsilon}\left(\frac{N}{m}\right) 
 & =\frac{N}{m}\left(1-\frac{1}{m}\left[\frac{\epsilon^{2}}{2}-\frac{m\epsilon^{2}}{2N}-\frac{\epsilon^{3}}{6m}+\frac{\epsilon^{3}}{2N}-\frac{m\epsilon^{3}}{3N^{2}}\right]\right).
\end{align*}
Define 
\[
\beta(N,m,\epsilon)=\frac{\epsilon^{2}}{2}-\frac{m\epsilon^{2}}{2N}-\frac{\epsilon^{3}}{6m}+\frac{\epsilon^{3}}{2N}-\frac{m\epsilon^{3}}{3N^{2}}
\]
and note that for any $N\geq 6$ and $m\leq N/2$ we have 
\[
\underline{\beta}(\epsilon):=\frac{\epsilon^{2}}{4}\leq\beta(N,m,\epsilon),
\]
which is clear upon noticing that  $\beta(N,m,\epsilon)$
as a function of $m$ on $[1,N/2]$ is concave and thus reaches its minimum in $1$ or $N/2$ (and this minimum is greater than $\epsilon^2/4$, provided $N\geq 6$). For any $x\geq N/m_{j+1}$ we
can check that 
\[
g_{N,\epsilon}(x)\leq\frac{g_{N,\epsilon}(N/m_{j+1})}{N/m_{j+1}}\times x
\]
by noticing that $g_{N,\epsilon}$ is concave and that $g_{N,\epsilon}(x)\leq x$ for $x\in [0,N]$. Hence for $k\geq0$ such that $u_{k-1}\geq N/m_{j+1}$,
we have 
\[
u_{k}\leq\left(1-\frac{1}{m_{j+1}}\underline{\beta}(\epsilon)\right)u_{k-1}.
\]
Now suppose that for some $k_{j}\geq0$ we have $u_{k_{j}}\leq N/m_{j}$.
Then let us find $K$ such that $u_{k_{j}+K}\leq N/m_{j+1}$. It is
sufficient to find $K$ such that 
\begin{align*}
 & \left(1-\frac{1}{m_{j+1}}\underline{\beta}(\epsilon)\right)^{K}\frac{N}{m_{j}}\leq\frac{N}{m_{j+1}}\\
 & \Leftrightarrow K\geq\log\frac{m_{j+1}}{m_{j}}\left(-\log\left(1-\frac{1}{m_{j+1}}\underline{\beta}(\epsilon)\right)\right)^{-1}.
\end{align*}
Finally by using 
\[
\forall x\in(0,1)\quad\frac{1}{x}-1\leq\frac{1}{-\log(1-x)}\leq\frac{1}{x}
\]
we conclude that $K$ defined as 
\[
K=\left\lceil \left(\log\frac{m_{j+1}}{m_{j}}\right)\frac{m_{j+1}}{\underline{\beta}(\epsilon)}\right\rceil 
\]
guarantees the inequality $u_{k_{j}+K}\leq N/m_{j+1}$. In other words
$(u_{k})_{k\geq0}$ needs less than $K$ steps to decrease from $N/m_{j}$
to $N/m_{j+1}$. Summing the terms between $k_{j}$ and $k_{j}+K$,
we obtain 
\begin{align*}
\sum_{k=k_{j}}^{k_{j}+K}u_{k} & \leq K\frac{N}{m_{j}}
  \leq\left[\left(\log\frac{m_{j+1}}{m_{j}}\right)\frac{m_{j+1}}{\underline{\beta}(\epsilon)}+1\right]\frac{N}{m_{j}}.
\end{align*}
Taking $m_{j}=2^{j}$ and $m_{j+1}=2^{j+1}$, we have $k_{j+1}\leq k_{j}+K$
and thus obtain 
\begin{align*}
\sum_{k=k_{j}}^{k_{j+1}}u_{k}\leq\sum_{k=k_{j}}^{k_{j}+K}u_{k} 
 & \leq\left[\left(\log2\right)\frac{2}{\underline{\beta}(\epsilon)}+\frac{1}{2^{j}}\right]N = C(\epsilon)N
\end{align*}
with $C(\epsilon)$ independent of $N$.
We have thus bounded the full sum by
\begin{align*}
\sum_{k\geq0}(u_{k}-1) & \leq\sum_{k=0}^{\sigma_{2}}(u_{k}-1)+\sum_{k\geq\sigma_{2}}(u_{k}-1)\\
 & \leq\left\lceil \frac{\log N}{\log2}\right\rceil C(\epsilon)N+\frac{2N}{\epsilon^{2}} \leq D(\epsilon) N \log N
\end{align*}
for some $D(\epsilon)$ independent of $N$.

\bibliographystyle{spbasic}      
\bibliography{AncestryCache}

\begin{thebibliography}{26}
\providecommand{\natexlab}[1]{#1}
\providecommand{\url}[1]{{#1}}
\providecommand{\urlprefix}{URL }
\expandafter\ifx\csname urlstyle\endcsname\relax
  \providecommand{\doi}[1]{DOI~\discretionary{}{}{}#1}\else
  \providecommand{\doi}{DOI~\discretionary{}{}{}\begingroup
  \urlstyle{rm}\Url}\fi
\providecommand{\eprint}[2][]{\url{#2}}

\bibitem[{Andrieu et~al(2010)Andrieu, Doucet, and
  Holenstein}]{andrieu:doucet:holenstein:2010}
Andrieu C, Doucet A, Holenstein R (2010) Particle {M}arkov chain {M}onte
  {C}arlo (with discussion). J Royal Statist Society Series B 72(4):357--385

\bibitem[{Capp{{\'e}} et~al(2005)Capp{{\'e}}, Moulines, and
  Ryd{{\'e}}n}]{cappe:ryden:2004}
Capp{{\'e}} O, Moulines E, Ryd{{\'e}}n T (2005) Inference in Hidden {M}arkov
  Models. Springer-Verlag, New York

\bibitem[{Carpenter et~al(1999)Carpenter, Clifford, and
  Fearnhead}]{CarClifFearn}
Carpenter J, Clifford P, Fearnhead P (1999) Improved particle filter for
  nonlinear problems. IEE Proc Radar, Sonar Navigation 146(1):2--7

\bibitem[{{Chopin} and {Singh}(2013)}]{Chopin:Singh:2013}
{Chopin} N, {Singh} SS (2013) {On the particle Gibbs sampler}. ArXiv e-prints
  \eprint{1304.1887}

\bibitem[{Chopin et~al(2013)Chopin, Jacob, and
  Papaspiliopoulos}]{chopin2012smc2}
Chopin N, Jacob P, Papaspiliopoulos O (2013) {SMC$^2$}: an efficient algorithm
  for sequential analysis of state space models. Journal of the Royal
  Statistical Society: Series B (Statistical Methodology) 75(3):397--426

\bibitem[{Del~Moral(2004)}]{delMoral:book}
Del~Moral P (2004) {F}eynman-{K}ac formulae. Springer

\bibitem[{Del~Moral and Doucet(2003)}]{del2003class}
Del~Moral P, Doucet A (2003) On a class of genealogical and interacting
  metropolis models. S{\'e}minaire de Probabilit{\'e}s XXXVII pp 415--446

\bibitem[{Del~Moral et~al(2009)Del~Moral, Miclo, Patras, and
  Rubenthaler}]{del2009convergence}
Del~Moral P, Miclo L, Patras F, Rubenthaler S (2009) The convergence to
  equilibrium of neutral genetic models. Stochastic Analysis and Applications
  28(1):123--143

\bibitem[{Douc et~al(2012)Douc, Moulines, and
  Olsson}]{Douc:Moulines:Olsson:2012}
Douc R, Moulines E, Olsson J (2012) Long-term stability of sequential {M}onte
  {C}arlo methods under verifiable conditions. ArXiv e-prints; to appear in
  Annals of Applied Probability \eprint{1203.6898}

\bibitem[{Doucet and Johansen(2011)}]{doucet2011tutorial}
Doucet A, Johansen A (2011) A tutorial on particle filtering and smoothing:
  Fifteen years later. In: Handbook of Nonlinear Filtering, Oxford, UK: Oxford
  University Press

\bibitem[{Doucet et~al(2001)Doucet, {{de Freitas}}, and
  Gordon}]{doucet:defreitas:gordon:2001}
Doucet A, {{de Freitas}} N, Gordon N (2001) Sequential {M}onte {C}arlo methods
  in practice. Springer-Verlag, New York

\bibitem[{Gordon et~al(1993)Gordon, Salmond, and
  Smith}]{gordon:salmon:smith:1993}
Gordon N, Salmond J, Smith A (1993) A novel approach to
  non-linear/non-{G}aussian {B}ayesian state estimation. IEEE Proceedings on
  Radar and Signal Processing 140:107--113

\bibitem[{van Handel(2009)}]{vanHandel:2009}
van Handel R (2009) Uniform time average consistency of monte carlo particle
  filters. Stochastic Processes and their Applications 119(11):3835 -- 3861

\bibitem[{Jones et~al(2010)Jones, Parslow, and Murray}]{jones2010bayesian}
Jones EM, Parslow J, Murray LM (2010) A {B}ayesian approach to state and
  parameter estimation in a phytoplankton-zooplankton model. Australian
  Meteorological and Oceanographic Journal 59:7--16

\bibitem[{Kitagawa(1998)}]{Kitagawa}
Kitagawa G (1998) A self-organizing state-space model. J American Statist Assoc
  93:1203--1215

\bibitem[{Lee et~al(2010)Lee, Yau, Giles, Doucet, and Holmes}]{lee2010}
Lee A, Yau C, Giles MB, Doucet A, Holmes CC (2010) On the utility of graphics
  cards to perform massively parallel simulation of advanced {M}onte {C}arlo
  methods. Journal of Computational and Graphical Statistics 19:769--789

\bibitem[{Lindsten et~al(2012)Lindsten, Jordan, and
  Sch{\"o}n}]{lindsten2012ancestor}
Lindsten F, Jordan M, Sch{\"o}n T (2012) Ancestor sampling for particle
  {G}ibbs. ArXiv e-prints \eprint{1210.6911}

\bibitem[{Liu and Chen(1998)}]{LiuChen}
Liu J, Chen R (1998) Sequential {M}onte {C}arlo methods for dynamic systems. J
  American Statist Assoc 93:1032--1044

\bibitem[{M{\"o}hle(2004)}]{mohle2004time}
M{\"o}hle M (2004) The time back to the most recent common ancestor in
  exchangeable population models. Advances in Applied Probability pp 78--97

\bibitem[{Murray(2013)}]{murray2013libbi}
Murray LM (2013) Bayesian state-space modelling on high-performance hardware
  using {LibBi}. ArXiv e-prints \eprint{1306.3277}

\bibitem[{{Murray} et~al(2012){Murray}, {Jones}, and
  {Parslow}}]{murray2012collapsed}
{Murray} LM, {Jones} EM, {Parslow} J (2012) On collapsed state-space models and
  the particle marginal {M}etropolis-{H}astings sampler. ArXiv e-prints
  \eprint{1202.6159}

\bibitem[{Murray et~al(2013)Murray, Lee, and Jacob}]{murray2013}
Murray LM, Lee A, Jacob PE (2013) Rethinking resampling in the particle filter
  on graphics processing units. ArXiv e-prints \eprint{1301.4019}

\bibitem[{Poyiadjis et~al(2011)Poyiadjis, Doucet, and
  Singh}]{poyiadjis2011particle}
Poyiadjis G, Doucet A, Singh S (2011) Particle approximations of the score and
  observed information matrix in state space models with application to
  parameter estimation. Biometrika 98(1):65--80

\bibitem[{Sengupta et~al(2008)Sengupta, Harris, and Garland}]{sengupta2008}
Sengupta S, Harris M, Garland M (2008) Efficient parallel scan algorithms for
  {GPU}s. Tech. Rep. NVR-2008-003, NVIDIA

\bibitem[{{W}ang et~al(2014){W}ang, {J}asra, and {De
  {Iorio}}}]{wang:jasra:2013}
{W}ang J, {J}asra A, {De {Iorio}} M (2014) Computational methods for a class of
  network models. Journal of Computational Biology

\bibitem[{{Whiteley}(2011)}]{whiteley2011stability}
{Whiteley} N (2011) {Stability properties of some particle filters}. ArXiv
  e-prints; to appear in Annals of Applied Probability \eprint{1109.6779}

\end{thebibliography}

\end{document}